# Competition between Ferrimagnetism and Magnetic Frustration in Zinc Substituted YBaFe$_4$O$_7$

Tapati Sarkar[*], V. Pralong, V. Caignaert and B. Raveau

*Laboratoire CRISMAT, UMR 6508 CNRS ENSICAEN,
6 bd Maréchal Juin, 14050 CAEN, France*

**Abstract**

The substitution of zinc for iron in YBaFe$_4$O$_7$ has allowed the oxide series YBaFe$_{4-x}$Zn$_x$O$_7$, with $0.40 \leq x \leq 1.50$, belonging to the "114" structural family to be synthesized. These oxides crystallize in the hexagonal symmetry (*P6$_3$mc*), as opposed to the cubic symmetry (*F-43m*) of YBaFe$_4$O$_7$. Importantly, the d.c. magnetization shows that the zinc substitution induces ferrimagnetism, in contrast to the spin glass behaviour of YBaFe$_4$O$_7$. Moreover, a.c. susceptibility measurements demonstrate that concomitantly these oxides exhibit a spin glass or a cluster glass behaviour, which increases at the expense of ferrimagnetism, as the zinc content is increased. This competition between ferrimagnetism and magnetic frustration is interpreted in terms of lifting of the geometric frustration, inducing the magnetic ordering, and of cationic disordering, which favours the glassy state.

* Corresponding author: Dr. Tapati Sarkar
e-mail:tapati.sarkar@ensicaen.fr
Fax: +33 2 31 95 16 00
Tel:  +33 2 31 45 26 32

**Introduction**

Transition metal oxides, involving a mixed valence of the transition element, present a great potential for the generation of strongly correlated electron systems. An extraordinary number of studies have been carried out in the last two decades on systems such as superconducting cuprates, colossal magnetoresistive manganites, and magnetic cobaltites, whose oxygen lattice exhibits a "square" symmetry derived from the perovskite structure. In contrast, the number of oxides with a triangular symmetry of the oxygen framework is much more limited, if one excludes the large family of spinels which exhibit exceptional ferrimagnetic properties and unique magnetic transitions [1 – 3] as for $Fe_3O_4$, but also magnetic frustration [4 – 5] due to the peculiar triangular geometry of their framework as for spinel ferrite thin films.

The recent discovery of the new series of cobaltites $(Ln,Ca)_1BaCo_4O_7$ [6 – 8] and ferrites $(Ln,Ca)_1BaFe_4O_7$ [9 – 11], termed "114" oxides, have opened up a new field for the investigation of strongly correlated electron systems. These oxides are closely related to spinels and barium hexaferrites, by their close packing of "$O_4$" and "$BaO_3$" layers. Importantly, they differ from all the other mixed valent transition metal oxides by the fact that cobalt or iron exists only in the tetrahedral coordination. Moreover, the symmetry of the structure may vary according to the temperature as well as the nature of the transition element, cobalt or iron, from hexagonal to cubic or orthorhombic. This change of symmetry seems to dramatically influence the magnetic properties, ranging from spin glass behaviour for cubic $LnBaFe_4O_7$ oxides [10 – 11] to ferrimagnetism for orthorhombic $CaBaCo_4O_7$ [8] and hexagonal $CaBaFe_4O_7$ [9] oxides. Very complex magnetic transitions are observed, as exemplified for $YBaCo_4O_7$, for which a spin glass behaviour was first reported around 66 K [7], whereas long range magnetic order was observed below $T_N = 110$ K [12], and a magnetic transition with short range correlations was revealed above $T_N$ [13]. Moreover, such magnetic transitions are often induced by structural transitions ($T_S$), as shown by the lifting of the magnetic frustration at $T_S$ in $LnBaCo_4O_7$ oxides [6, 7, 14].

The remarkable feature of the "114" ferrites is the existence of the cubic symmetry of $LnBaFe_4O_7$ oxides [10 – 11] with Ln = Dy, Ho, Er, Yb, Lu, which has never been observed in the cobaltite series, whereas the hexagonal symmetry is observed for both types of oxides, $LnBaCo_4O_7$ [6 – 7] oxides whatever Ln, as well as $CaBaFe_4O_7$ [9] and $LnBaFe_4O_7$ [11] with Ln = Tb, Gd. Understanding the relative stability of these two structural types of ferrites is of great importance since it appears that it governs their magnetic properties, leading to spin glass behaviour and ferrimagnetism for cubic and hexagonal $LnBaFe_4O_7$ oxides respectively [11]. In fact, the two structures are closely related, i.e. built up of an ordered 1:1 stacking of identical tetrahedral layers, i.e. kagomé (K) and triangular (T) layers. The cubic structure (**Fig. 1(a)**) can then be deduced from the hexagonal one (**Fig. 1 (b)**) by translation of one T layer out of two by $\frac{\vec{a}}{2}$. Bearing in mind that the structural transition from cubic to hexagonal symmetry is very sensitive to geometric factors, as shown from the effect of the size of $Ln^{3+}$ cations [11], we have explored the effects of cationic substitutions upon the structural and magnetic properties of the cubic spin glass $YBaFe_4O_7$. In this study, we show that the substitution of $Zn^{2+}$ for $Fe^{2+}$ in this phase allows the hexagonal symmetry to be stabilized for the series $YBaFe_{4-x}Zn_xO_7$ with $0.40 \leq x \leq 1.50$. Moreover, it is observed that this substitution by a diamagnetic cation induces a ferrimagnetic ground state, in agreement with the change of symmetry. However, since $Zn^{2+}$ is a diamagnetic cation, it does not participate in the magnetic interactions, but rather dilutes the magnetic sublattice and introduces cationic disordering, which in turn is responsible for the appearance of magnetic frustration. The combined measurements of a.c. and d.c. magnetic susceptibility throw light on the competition between ferrimagnetic interactions and magnetic frustration in this series, suggesting a possible phase separation.

**Experimental**

Phase-pure samples of $YBaFe_{4-x}Zn_xO_7$ [x = 0.40 – 1.50] were prepared by solid state reaction technique. The precursors used were $Y_2O_3$, $BaFe_2O_4$, ZnO, $Fe_2O_3$ and metallic Fe powder. First, the precursor $BaFe_2O_4$ was prepared from a stoichiometric mixture of $BaCO_3$ and $Fe_2O_3$ annealed at 1200°C for 12 hrs in air. In a second step, a stoichiometric mixture of $Y_2O_3$, $BaFe_2O_4$, ZnO, $Fe_2O_3$ and metallic Fe powder was intimately ground and pressed in the form of rectangular bars. The bars were then kept in an alumina finger, sealed in silica tubes under vacuum and annealed at 1100°C for 12 hrs. Finally, the samples were cooled to room temperature at a rate of 200°C / hr.

The X-ray diffraction patterns were registered with a Panalytical X'Pert Pro diffractometer under a continuous scanning mode in the 2θ range 10° - 120° and step size Δ2θ = 0.017°. The d.c. magnetization measurements were performed using a superconducting quantum interference device (SQUID) magnetometer with variable temperature cryostat (Quantum Design, San Diego, USA). The a.c. susceptibility, $\chi_{ac}(T)$ was measured with a PPMS from Quantum Design with the frequency ranging from 10 Hz to 10 kHz ($H_{dc}$ = 0 Oe and $H_{ac}$ = 10 Oe). All the magnetic properties were registered on dense ceramic bars of dimensions ~ 4 × 2 × 2 mm$^3$.

**Results and discussion**

For the above synthesis conditions, a monophasic system was obtained for YBaFe$_{4-x}$Zn$_x$O$_7$ with 0.40 ≤ x ≤ 1.50. The EDS analysis of the samples (**Table 1**) allowed the cationic compositions to be confirmed.

*Structural characterization*

The X-ray diffraction (XRD) patterns show that all the samples with 0.40 ≤ x ≤ 1.50 exhibit the hexagonal symmetry, as exemplified for the two extreme members of the series, x = 0.40 and x = 1.50 (**Fig. 2**), whose fits have been achieved by Rietveld analysis using the FULLPROF refinement program [15]. The extracted cell parameters (**Table 1**) show that "*a*" increases slightly as x increases, whereas "*c*" decreases so that the cell volume does not vary significantly, in agreement with the size of Zn$^{2+}$ ($r_{Zn^{2+}}$ ≈ 0.60 Å) being very similar to that of Fe$^{2+}$ ($r_{Fe^{2+}}$ ≈ 0.63 Å). Thus, the stabilization of the hexagonal phase with respect to the cubic one is probably not due to the smaller size of the Zn$^{2+}$ cation. The rather different distortion of the ZnO$_4$ tetrahedra compared to the Fe$^{II}$O$_4$ tetrahedra may be at the origin of this structural transition allowing a different tuning of the kagomé and triangular layers in the hexagonal phase compared to the cubic one.

A preferential occupancy of one of the two different kinds of sites by Zn$^{2+}$ in the hexagonal structure is possible, but cannot be detected here from XRD data due to the very close values of the scattering factors of Zn and Fe. In any case, the *c* / *a* ratio of the hexagonal cell of the series YBaFe$_{4-x}$Zn$_x$O$_7$ (**Table 1**) clearly shows that the introduction of zinc in the structure tends to destroy the close packing of the "BaO$_3$" and "O$_4$" layers, leading to a *c* / *a* value significantly larger than the ideal value for a perfect close packed structure (1.6333), and in the same way ruling out a possible cubic close packed symmetry. Note, however, that the decrease of *c* / *a* ratio as the zinc content increases is not yet understood.

It is also quite remarkable that attempts to synthesize phase pure samples for 0 < x < 0.40 were unsuccessful, instead leading to a mixture of the cubic and hexagonal oxides. This observation is important, since it demonstrates that in this region of cross-over from cubic to hexagonal symmetry, no cubic solid solution can be obtained, confirming that Zn$^{2+}$ cannot accommodate the unique tetrahedral site of the cubic structure of YBaFe$_4$O$_7$, most likely due to the different distortion of ZnO$_4$ tetrahedra compared to FeO$_4$ tetrahedra.

*D. C. magnetization studies*

The temperature dependence of d.c. magnetization was registered according to the standard zero field cooled (ZFC) and field cooled (FC) procedures. A magnetic field of 0.3 T was applied during the measurements. The measurements were done in a temperature range of 5 K to 400 K. The $M_{ZFC}(T)$ and $M_{FC}(T)$ curves of all the samples are shown in **Fig. 3**. The highest value of the magnetization was observed for the sample with the lowest zinc content, x = 0.40. For this sample, the magnetization increases sharply as the sample is cooled below 150 K, as is seen from the FC and ZFC curves. The FC curve shows saturation at about 0.78 $\mu_B$/f.u. at 5 K, with the typical shape of ferro (or ferri) magnetic behaviour, whereas the ZFC curve exhibits a maximum at ~ 80 K reflecting the existence of strong irreversibility. This behaviour is in contrast to the spin glass behaviour of $YBaFe_4O_7$ [10]. Rather, it is comparable to $CaBaFe_4O_7$ which was found to be ferrimagnetic with a hexagonal symmetry. Nevertheless, the $T_C$ value of $YBaFe_{3.6}Zn_{0.4}O_7$ is significantly smaller than the value obtained for $CaBaFe_4O_7$ (270 K). Also, the Curie-Weiss temperature ($\theta_{CW}$) for $YBaFe_{3.6}Zn_{0.4}O_7$, obtained by extrapolation of the linear $\chi^{-1}(T)$ region, is 63.7 K. This is substantially lower than the value that was obtained for $CaBaFe_4O_7$ (268.3 K). This suggests that the dilution introduced by the diamagnetic cation $Zn^{2+}$ weakens the ferrimagnetic interaction in this system, and the final ground state of the oxide may not be a simple ferrimagnet, but a more complex system involving magnetic frustration.

With an increase in the substitution level (x), the value of the magnetization attained by the sample at the lowest measured temperature of 5 K decreases monotonically. (This decrease is shown in the inset of **Fig. 3**.) This feature is again a reflection of the fact that the $Zn^{2+}$ ions do not carry any magnetic moment, and thus, serve to dilute the ferrimagnetic coupling. This dilution of the ferrimagnetic interaction is also seen in the fact that the temperature below which the magnetization starts rising, as well as the sharpness of the transition keeps decreasing with an increase in the zinc content (x) (**Fig. 3**). This has been shown quantitatively in **Fig. 4**, where we have plotted the first derivative of magnetization (dM/dT) versus temperature curves. The inflection points of the curves ($T_{inf}$) can be roughly considered to be the temperatures at which the paramagnetic to ferrimagnetic transition takes place. (Later on we will show that the ground state obtained at low temperature in these oxides is not a simple ferrimagnet. Instead, it is rather complex, as will become clear from the a.c. magnetic susceptibility measurements, which we detail in the next section). Inset (a) of **Fig. 4** shows the monotonic decrease of $T_{inf}$ with increase in the substitution level (x).

In inset (b) of **Fig. 4**, we show the variation of $\Delta T_{1/2}$ with the doping concentration (x). $\Delta T_{1/2}$ is the full width at half maximum (F.W.H.M.) of the dM/dT vs T curves. It gives a measure of the width of the transition (a lower value of $\Delta T_{1/2}$ indicates a sharper transition). As is seen in the figure, the width of the transition shows a systematic increase with an increase in the substitution level (x), thereby indicating the dilution of the ferrimagnetic interactions.

In view of the observations on the irreversibility of the FC and ZFC M(T) curves seen at low temperature, detailed investigations were carried out on the d.c. magnetization M(H) curves of the various samples at 5 K. The isothermal magnetization curves recorded at 5 K (**Fig. 5**) show that the x = 0.40 sample exhibits the largest hysteresis loop i.e. the largest values of the coercive field ($H_C \sim 1$ T) and of the remanent magnetization ($M_r \sim 0.8$ $\mu_B$/f.u.), indicating that $YBaFe_{3.6}Zn_{0.4}O_7$ is, like $CaBaFe_4O_7$, a hard ferrimagnet. Importantly, the shape of the M(H) loop of this phase is much smoother than for $CaBaFe_4O_7$, and rather similar to that observed for $CaBaFe_{4-x}Li_xO_7$ oxides, that were shown to be glassy ferrimagnets [16]. These samples show similar magnetic hysteresis loops, characterized by a lack of magnetic saturation. Moreover, one observes that the coercive field $H_C$ (inset (a) of **Fig. 5**) and the remanent magnetization $M_r$ (inset (b) of **Fig. 5**) exhibit a monotonic decrease as the substitution level increases.

The lack of magnetic saturation in these samples can be quantified by taking the ratio of the value of the magnetization attained at a field of 5 T and $M_r$ [M (H = 5 T) / $M_r$]. This is shown in **Fig. 6 (a)**. As can be seen from the figure, the quantity M (H = 5 T) / $M_r$ increases by nearly 5 times as the doping concentration is increased from x = 0.40 to x = 1.50. This again indicates the strong weakening of the ferrimagnetic interaction with increasing zinc concentration. The spontaneous magnetization (M at H = 0, as read out from the virgin curve) shows a monotonic decrease with an increase in the substitution level (x) (**Fig. 6 (b)**). As x increases from 0.40 to 1.50, the spontaneous magnetization decreases by 4 orders.

Thus, these results suggest that in these hexagonal oxides, ferrimagnetism competes with magnetic frustration due to the triangular geometry of the kagomé layers. Such a statement is also supported by the recent study of the isostructural "114" hexagonal cobaltite $YBaCo_4O_7$ [13] which shows that 1 D magnetic ordering competes with 2 D magnetic frustration.

*A. C. magnetic susceptibility studies*

The measurements of the a.c. magnetic susceptibility $\chi'_{ac}(T,f)$ were performed in zero magnetic field ($H_{dc} = 0$) at different frequencies ranging from 10 Hz to 10000 Hz, using a PPMS facility. The amplitude of the a.c. magnetic field was ~ 10 Oe. The evolution of the real part $\chi'(T)$ and the imaginary part $\chi''(T)$ of the susceptibility versus temperature at different frequencies for the two end members of the series, x = 0.40 (**Fig. 7**) and x = 1.50 (**Fig. 8**) sheds light on the complex nature of the magnetic ground state of these oxides. The $\chi'(T)$ curves (**Fig. 7 (a)** and **Fig. 8 (a)**) exhibit a maximum at $T_{max(a.c.)}$ which is slightly higher than the peak values obtained from d.c. magnetic measurements ($T_{max(d.c.)}$) i.e. $T_{max(a.c.)} = 105$ K at 10 Hz (**Fig. 7 (a)** and **Table 2**) against $T_{max(d.c.)} = 83$ K (**Fig. 3**) for the x = 0.40 sample, and $T_{max(a.c.)} = 45$ K (**Fig. 8 (a)** and **Table 2**) against $T_{max(d.c.)} = 36$ K for the x = 1.50 sample. Both samples show frequency dependent peaks in the a.c. susceptibility measurements i.e. as the frequency is increased, $T_{max(a.c.)}$ shifts to higher values associated with a decrease in the magnitude of $\chi'_{ac}$. This behaviour strongly suggests a spin glass [17] or superparamagnetic behaviour, which will be discussed below. The peak observed in the real part (in-phase component) of the magnetic susceptibility ($\chi'_{ac}(T)$) at a frequency of 10 Hz decreases from 105 K to 45 K as the doping concentration is increased from x = 0.40 to x = 1.50.

The plot of the imaginary part of the a.c. susceptibility (**Fig. 7 (b)** and **Fig. 8 (b)**) show that the maximum of $\chi''_{ac}(T)$ appears at a lower temperature than the maximum of $\chi'_{ac}(T)$, and is also frequency dependent. We note here that for the x = 0.40 sample, the magnitude of the peak in $\chi''_{ac}$ is largest for the lowest frequency measured (11 Hz). On the other hand, for the x = 1.5 sample, the magnitude of the peak in $\chi''_{ac}$ is largest for the highest frequency measured (10 kHz). This indicates that the inverse mean relaxation time of the spin system for the x = 0.40 sample is lower than that for the x = 1.50 sample. This is in accordance with the values of the spin relaxation time $\tau_0$ obtained later from the fitting of the a.c. susceptibility data ($\tau_{0(x = 0.40)} > \tau_{0(x = 1.50)}$) (see **Table 2).**

To check the existence of a spin glass behaviour or of cluster glass behaviour, we have analyzed the frequency dependence of the peak in $\chi'_{ac}(T)$ using the method previously developed by Bréard et.al. [18], starting from the dynamic scaling theory [17] which predicts a power law of the form $\tau = \tau_0 \left( \frac{T_f - T_{SG}}{T_{SG}} \right)^{-z\nu}$, where, $\tau_0$ is the shortest relaxation time available to the system, $T_{SG}$ is the underlying spin-glass transition temperature determined by the interactions in the system, $z$ is the dynamic critical exponent and $\nu$ is the critical exponent of the correlation length. The actual fittings were done using the equivalent form of the power law: $\ln \tau = \ln \tau_0 - z\nu \ln\left(\frac{T_f - T_{SG}}{T_{SG}}\right)$. The remarkable linearity obtained for the plots of $\ln \tau$ versus $\ln\left(\frac{T_f - T_{SG}}{T_{SG}}\right)$ using the fit parameters, shown in **Fig. 9**, indicate that these samples well obey the behaviour expected for a spin glass like system. Similar fittings were performed for all the other samples of the series, and the extracted parameters of this study have been listed in **Table 2**.

The parameter $p = \frac{\frac{\Delta T_f}{T_f}}{\Delta \log f}$ is known to vary from 0.004 to 0.02 for canonical spin glass systems. The value of $p$ that we obtain for the two end members of the series under investigation are $p = 0.015$ (for the x = 0.40 sample) and $p = 0.020$ (for the x = 1.50 sample). [Note: For superparamagnetic materials, the value of $p$ is higher (~ 0.1), and so we discard the possibility of these oxides being superparamagnetic.] In spite of the fact that the value of $p$ is quite similar for the two end members of the series, the fact that the shape of the a.c. susceptibility versus temperature is quite different in the two cases gets reflected in the values of the parameters $\tau_0$ and $z\nu$. In fact, for the lower zinc concentrations (x = 0.40 to x = 0.90), the spin relaxation time $\tau_0$ lies in the range of values typical of canonical spin glasses ($10^{-10} - 10^{-12}$ sec) [19 – 21], and the $z\nu$ values lie between 6 and 10. However, as the zinc concentration is increased beyond x = 1, the $\tau_0$ values are much smaller, and the $z\nu$ values obtained are above 10. Thus, the power law model provides a satisfactory description of our a.c. susceptibility data.

**Conclusion**

The present study of the zinc substituted "114" oxides $YBaFe_{4-x}Zn_xO_7$ shows that the substitution of zinc for iron in the $YBaFe_4O_7$ oxide induces a ferrimagnetic behaviour, in spite of the diamagnetic character of $Zn^{2+}$, but that these magnetic interactions compete with a spin glass or a cluster glass behaviour as the zinc content is further increased. In fact, the shape of the M-T curve obtained for this series of oxides (sharp decrease in the ZFC curve at lower temperatures with saturation of FC curve) is very similar to that reported for various alloys showing mictomagnetism[22,23,24], i.e. coexistence of frustrated spin clusters and ferrimagnetic spin regions. This competition between ferrimagnetism and magnetic frustration can be understood by considering the structure, or more exactly the iron lattice of these oxides. The cubic structure of YBaFe$_4$O$_7$ (**Fig. 1 (a)**) forms a tetrahedral [Fe$_4$]$_\infty$ lattice of Fe$_4$ tetrahedra sharing apices (**Fig. 10 (a)**), very similar to the [Ln$_4$]$_\infty$ lattice observed in pyrochlores. As previously shown for these compounds, this tetrahedral framework exhibits a complete 3 D geometric frustration and consequently, one observes, like for the pyrochlores [25], a spin glass behaviour [10]. As soon as a part of Zn$^{2+}$ is substituted for Fe$^{2+}$, i.e., for YBaFe$_{3.6}$Zn$_{0.4}$O$_7$ (x = 0.40), the structure becomes hexagonal (**Fig. 1 (b)**), forming a different [Fe$_8$]$_\infty$ lattice (**Fig. 10 (b)**), built up of rows of corner – sharing "Fe$_5$" trigonal bipyramids running along "*c*", and interconnected through "Fe$_3$" triangles in the "*a b*" plane. As a consequence, the zinc substitution induces a lifting of the geometric frustration, favouring the appearance of magnetic ordering, as previously observed for hexagonal CaBaFe$_4$O$_7$ which is known to be ferrimagnetic [9]. In fact, such a hexagonal framework could be regarded as composed of ferri (or ferro) magnetic rows running along "*c*" formed by the "Fe$_5$" bipyramids, whereas in the "*a b*" plane, the triangular geometry of the iron lattice is maintained so that a bidimensional magnetic frustration is still possible, and may compete with the 1 D magnetic ordering along "*c*". However, the presence of zinc on the iron sites tends to weaken the magnetic interactions due to its diamagnetic character by dilution effect, and the cationic disordering on the Fe / Zn sites favours the spin glass or cluster glass behaviour. As a consequence the ferrimagnetic behaviour decreases at the benefit of the spin glass or cluster glass behaviour as the zinc content increases beyond x = 0.40.

A neutron diffraction study of these glassy ferrimagnets will be necessary to determine the exact distribution of zinc on the two different sites of the structure and to understand the complex nature of the magnetic states that arise from the competition between ferrimagnetism and magnetic frustration. The possibility of phase separation should also be considered.

**Acknowledgements**

We gratefully acknowledge Dr. Vincent Hardy, CRISMAT, ENSICAEN, France for helpful discussions. We also acknowledge the CNRS and the Conseil Regional of Basse Normandie for financial support in the frame of Emergence Program.

**Table captions**

**Table 1**: Refined crystal structure parameters as obtained from the Rietveld refinement of X-ray powder diffraction data.

**Table 2**: Parameters extracted from fitting of a.c. magnetic susceptibility data.

**Figure captions**

**Figure 1:** Relative positions of the T layers in (a) cubic $YBaFe_4O_7$ and (b) hexagonal $CaBaFe_4O_7$ (adapted from Ref. 10). For details, see text.

**Figure 2:** X-ray diffraction patterns along with the fits for (a) $YBaFe_{3.6}Zn_{0.4}O_7$ and (b) $YBaFe_{2.5}Zn_{1.5}O_7$.

**Figure 3:** $M_{ZFC}(T)$ and $M_{FC}(T)$ curves of $YBaFe_{4-x}Zn_xO_7$ measured at H = 0.3 T. The empty symbols are for $M_{ZFC}(T)$ and the solid symbols are for $M_{FC}(T)$. The inset shows the variation of $M_{FC}$ at T = 5 K with the substitution level (x).

**Figure 4:** dM/dT vs T for $YBaFe_{4-x}Zn_xO_7$. Inset (a) shows the variation of $T_{inf}$ with x, inset (b) shows the variation of $\Delta T_{1/2}$ with x (see text for details).

**Figure 5:** M (H) for $YBaFe_{4-x}Zn_xO_7$ at T = 5 K. Inset (a) shows the variation of $H_C$ with x, inset (b) shows the variation of $M_r$ with x.

**Figure 6:** (a) M (H = 5 T) / $M_r$ vs x, (b) M (H = 0 T) vs x for $YBaFe_{4-x}Zn_xO_7$.

**Figure 7:** (a) Real (in-phase) and (b) Imaginary (out-of-phase) component of a.c. susceptibilities for $YBaFe_{3.6}Zn_{0.4}O_7$ as a function of temperature measured using a frequency range 10 Hz – 10 kHz.

**Figure 8:** (a) Real (in-phase) and (b) Imaginary (out-of-phase) component of a.c. susceptibilities for $YBaFe_{2.5}Zn_{1.5}O_7$ as a function of temperature measured using a frequency range 10 Hz – 10 kHz.

**Figure 9:** Plot of $\ln \tau$ vs $\ln\left(\dfrac{T_f - T_{SG}}{T_{SG}}\right)$ for (a) $YBaFe_{3.6}Zn_{0.4}O_7$ and (b) $YBaFe_{2.5}Zn_{1.5}O_7$ (for details, see text).

**Figure 10:** Schematic representation of (a) cubic $YBaFe_4O_7$ and (b) hexagonal $CaBaFe_4O_7$ (adapted from Ref. 11). For details, see text.

**Table 1**

| Doping concentration (x) | Crystal system (Space group) | Unit cell parameters | | $c/a$ | Reliability factor ($R_F$) | x as obtained from EDS analysis |
|---|---|---|---|---|---|---|
| | | $a$ (Å) | $c$ (Å) | | | |
| 0.40 | Hexagonal ($P6_3mc$) | 6.317(2) | 10.380(2) | 1.6432 | 3.11 | 0.36(0.02) |
| 0.50 | Hexagonal ($P6_3mc$) | 6.319(2) | 10.379(2) | 1.6425 | 8.31 | 0.53(0.03) |
| 0.60 | Hexagonal ($P6_3mc$) | 6.321(1) | 10.377(1) | 1.6417 | 5.30 | 0.65(0.04) |
| 0.80 | Hexagonal ($P6_3mc$) | 6.321(2) | 10.371(2) | 1.6407 | 9.28 | 0.80(0.04) |
| 0.90 | Hexagonal ($P6_3mc$) | 6.323(1) | 10.367(1) | 1.6396 | 7.97 | 0.90(0.04) |
| 1.25 | Hexagonal ($P6_3mc$) | 6.325(3) | 10.355(3) | 1.6372 | 8.31 | 1.27(0.03) |
| 1.50 | Hexagonal ($P6_3mc$) | 6.328(1) | 10.348(1) | 1.6353 | 8.24 | 1.56(0.03) |

**Table 2**

| Doping concentration (x) | $T_{max}$ from $\chi'(T)$ at $f =$ 10 Hz (K) | p | Spin relaxation time $\tau_0$ (sec) | Spin glass transition temperature $T_{SG}$ (K) | zν |
|---|---|---|---|---|---|
| 0.40 | 104.5 | 0.015 | $(1.95 \pm 0.04) \times 10^{-10}$ | 100.44 (7) | 6.2 (1) |
| 0.50 | 95.4 | 0.014 | $(8.14 \pm 0.04) \times 10^{-11}$ | 92.30 (3) | 6.6 (2) |
| 0.60 | 91.8 | 0.013 | $(3.09 \pm 0.08) \times 10^{-11}$ | 88.04 (2) | 6.9 (2) |
| 0.80 | 79.8 | 0.011 | $(1.11 \pm 0.04) \times 10^{-11}$ | 76.05 (2) | 7.3 (1) |
| 0.90 | 75.8 | 0.011 | $(2.93 \pm 0.01) \times 10^{-12}$ | 72.96 (6) | 7.4 (1) |
| 1.25 | 56.0 | 0.015 | $(8.75 \pm 0.09) \times 10^{-13}$ | 54.26 (2) | 10.8 (2) |
| 1.50 | 44.8 | 0.020 | $(3.75 \pm 0.04) \times 10^{-15}$ | 40.63 (6) | 13.6 (4) |

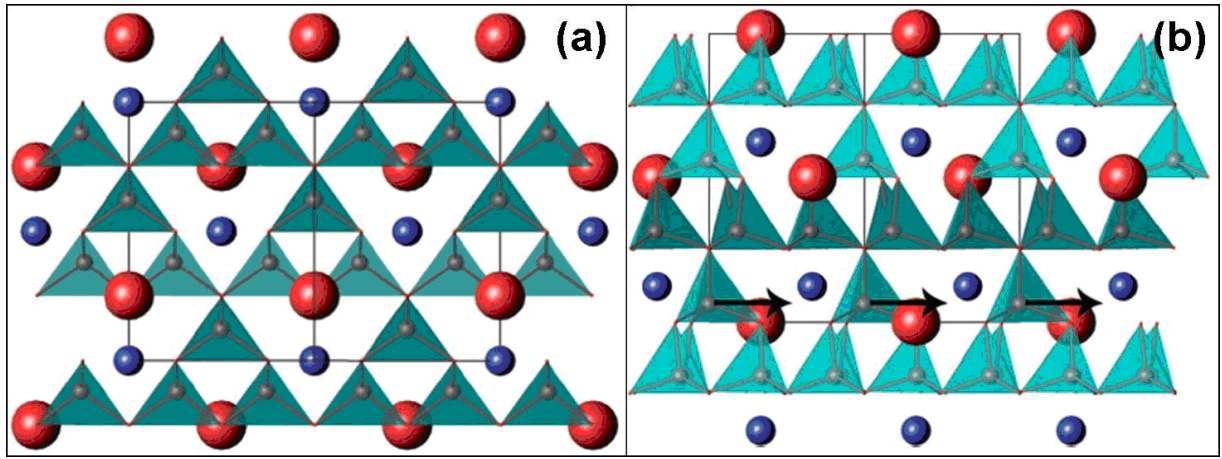

Figure 1

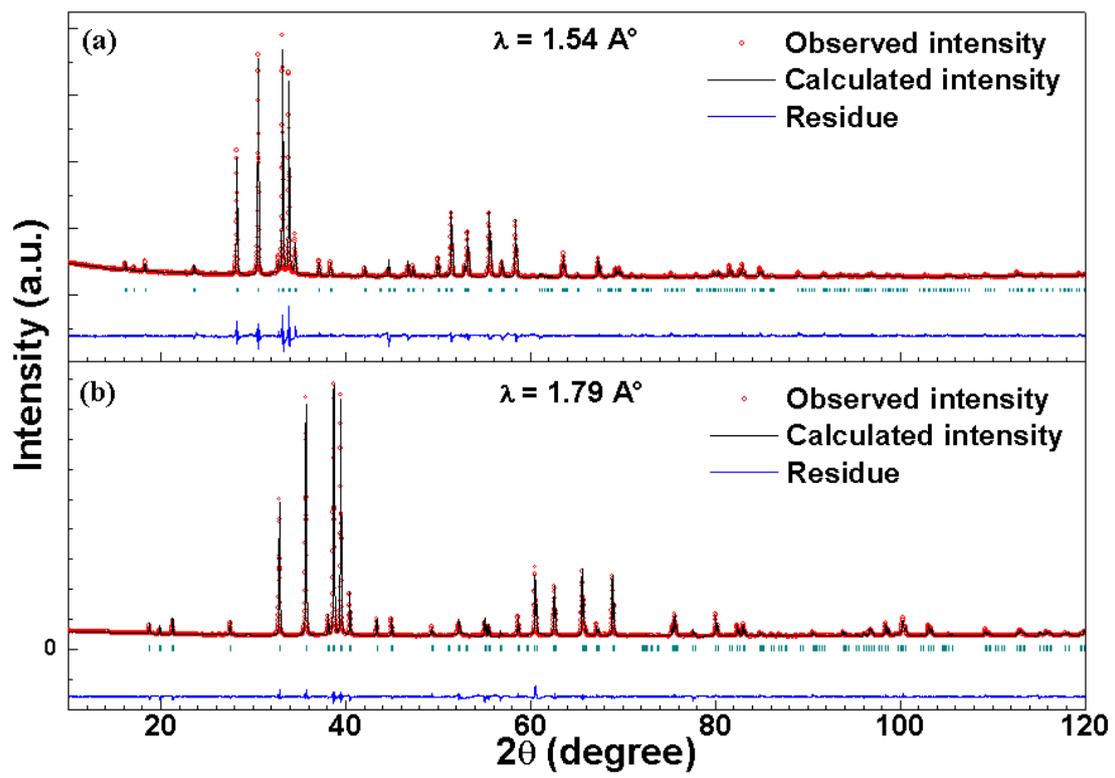

Figure 2

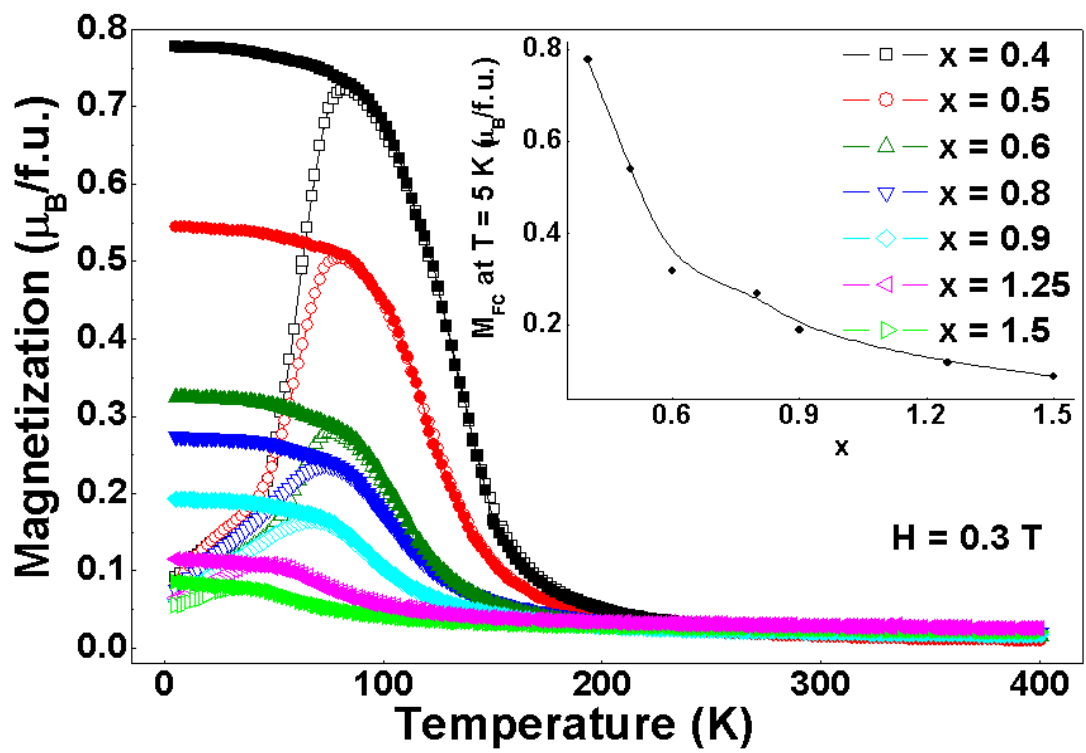

Figure 3

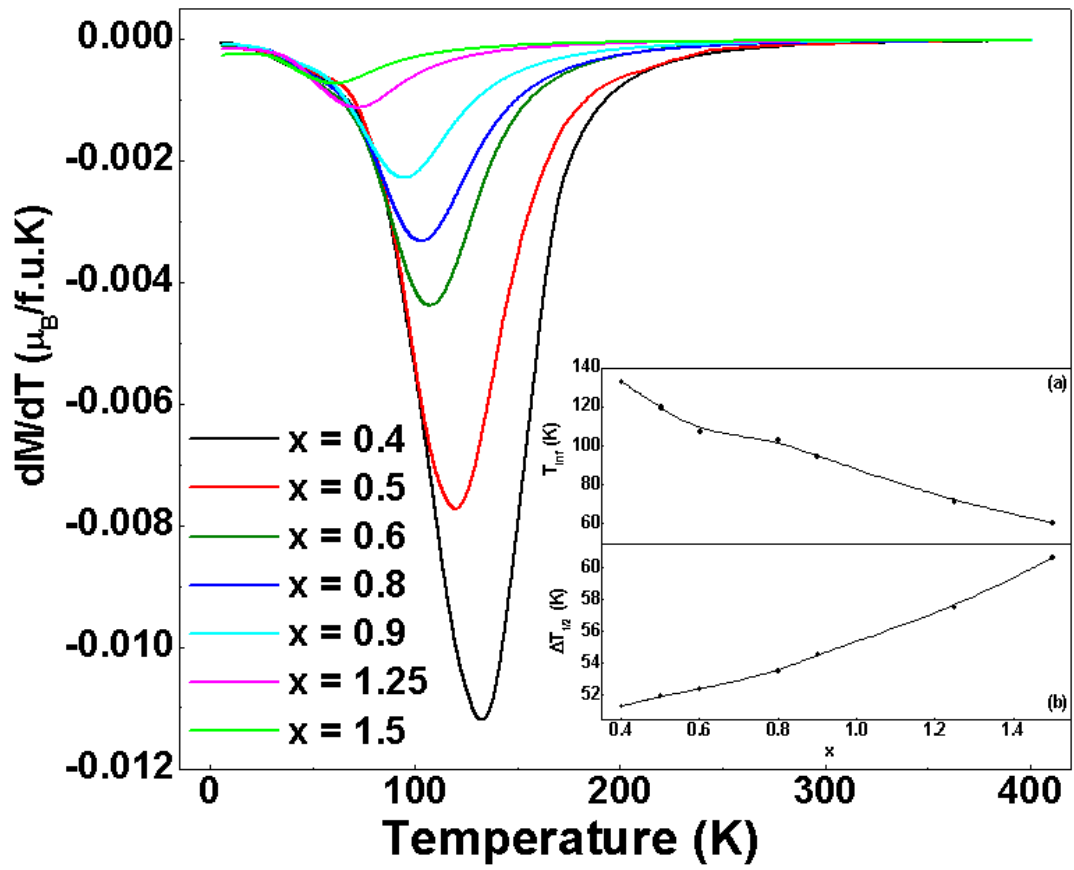

Figure 4

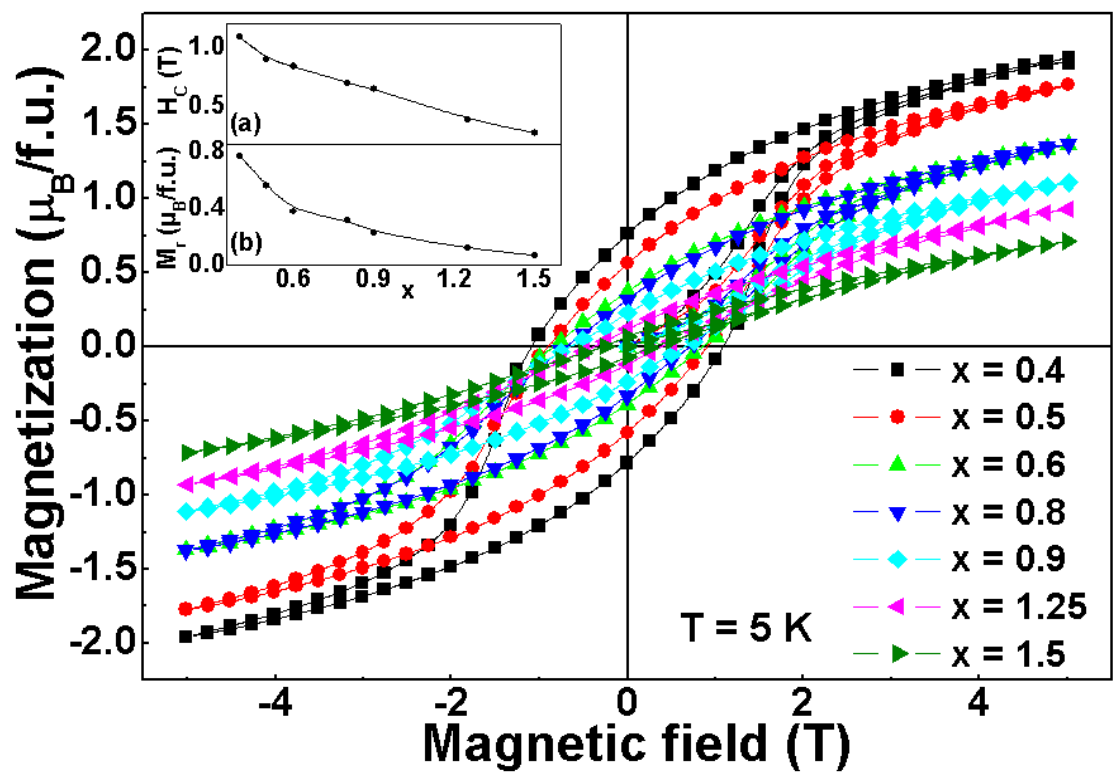

Figure 5

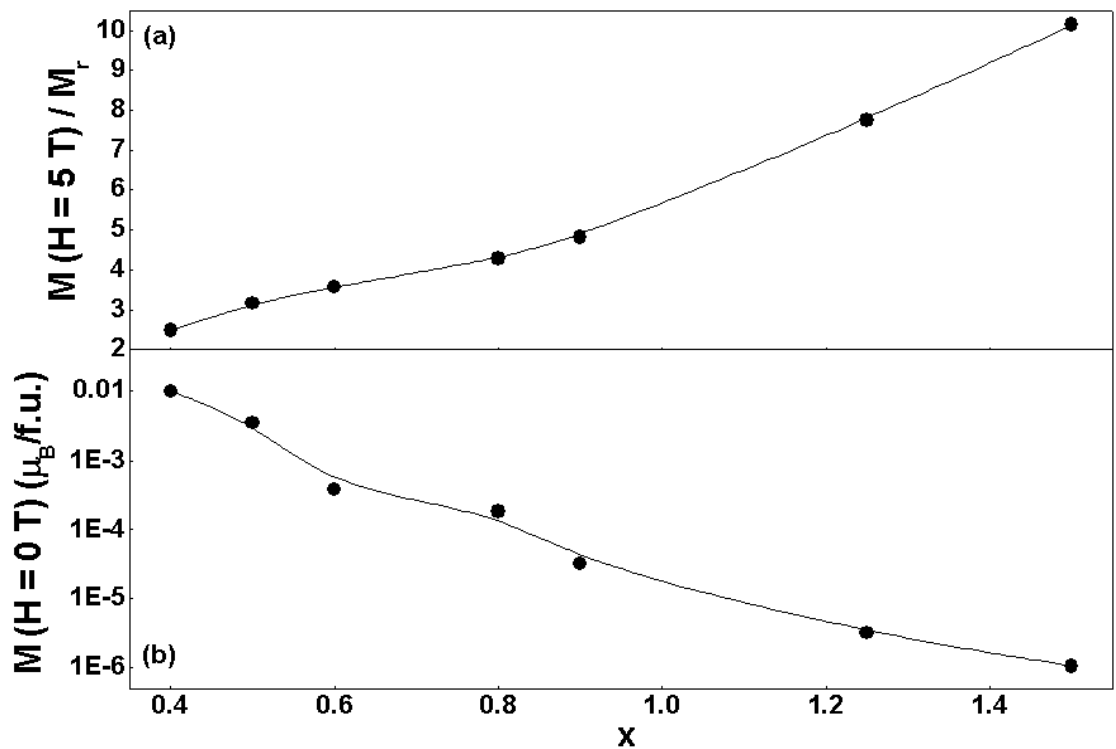

Figure 6

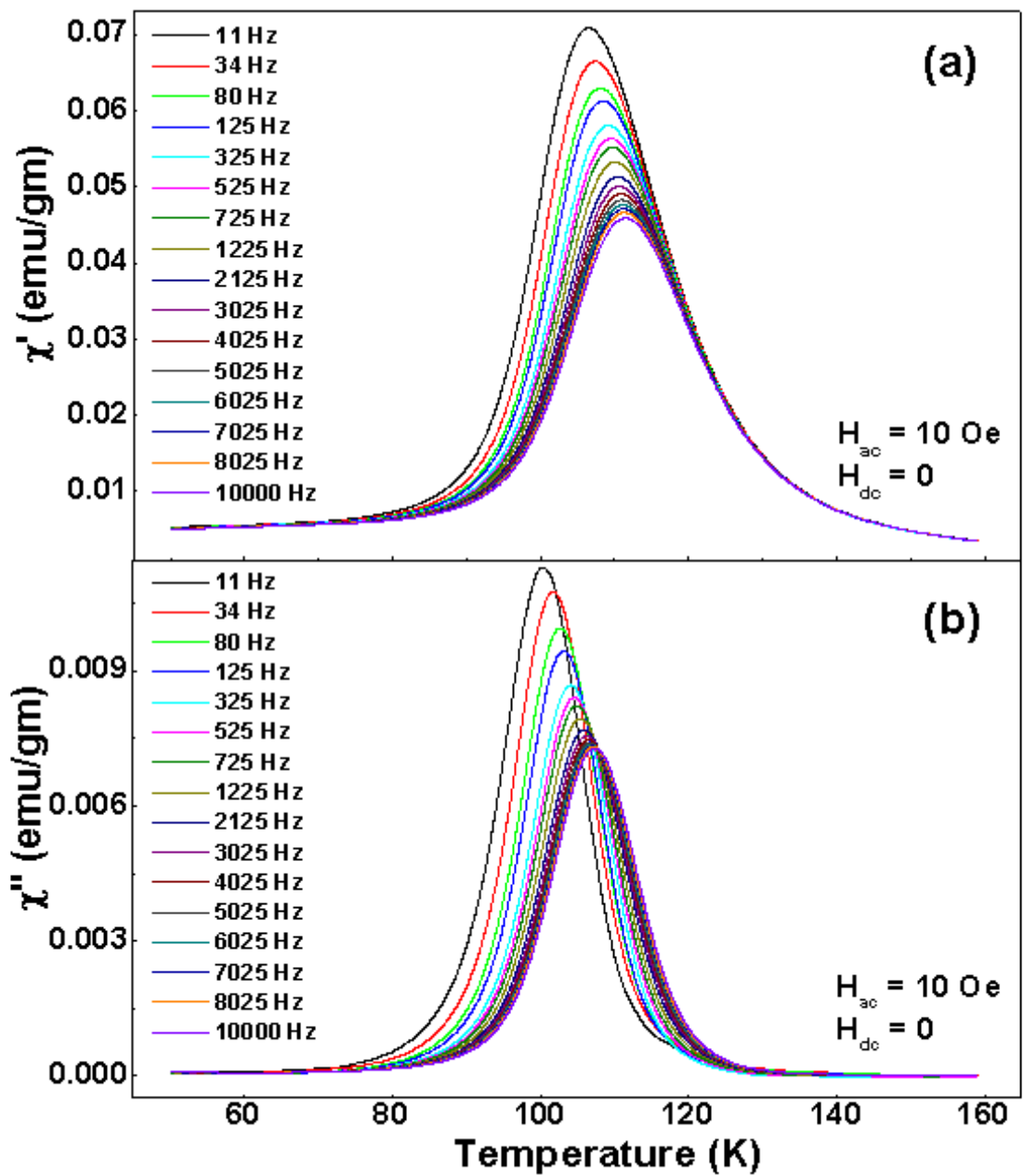

Figure 7

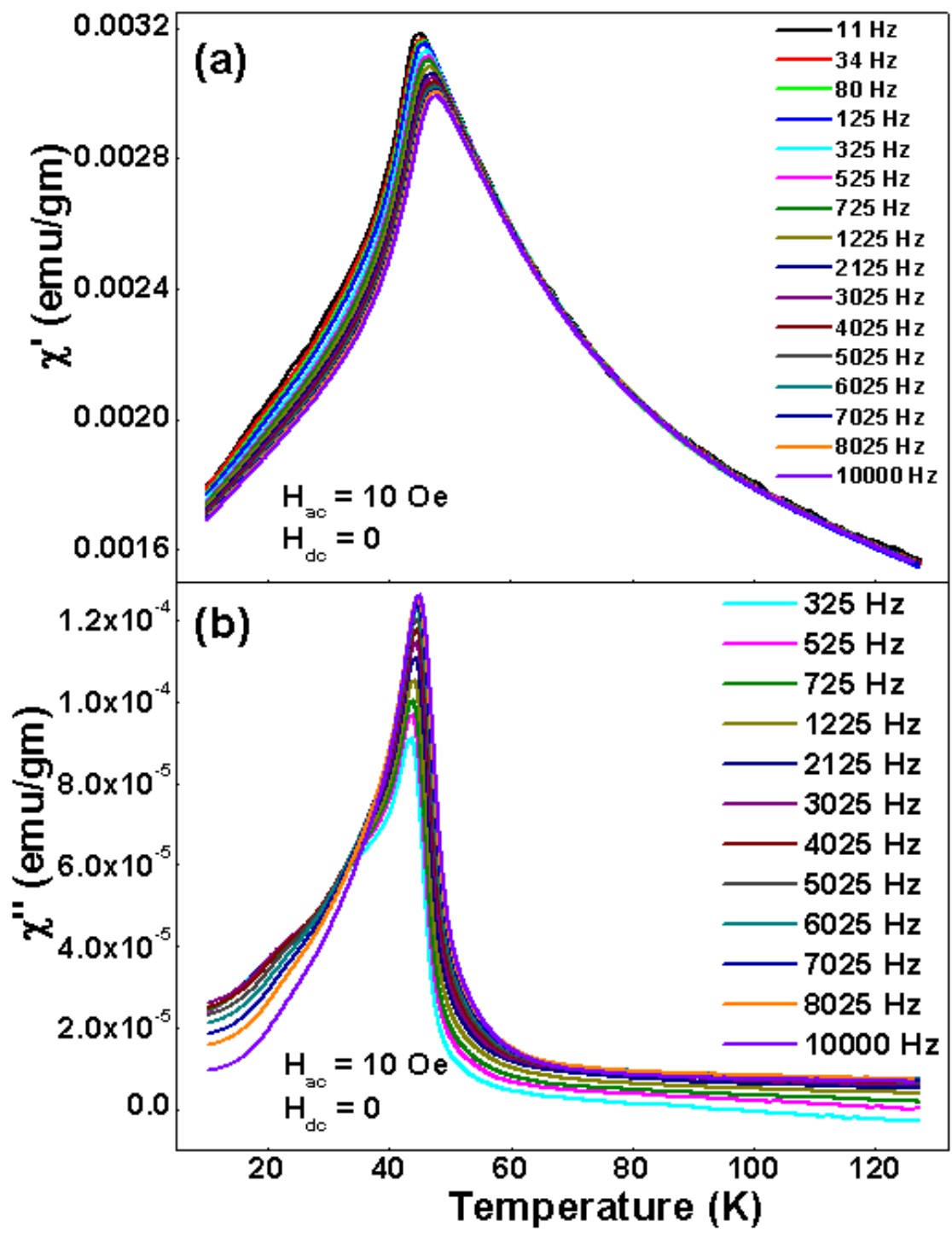

Figure 8

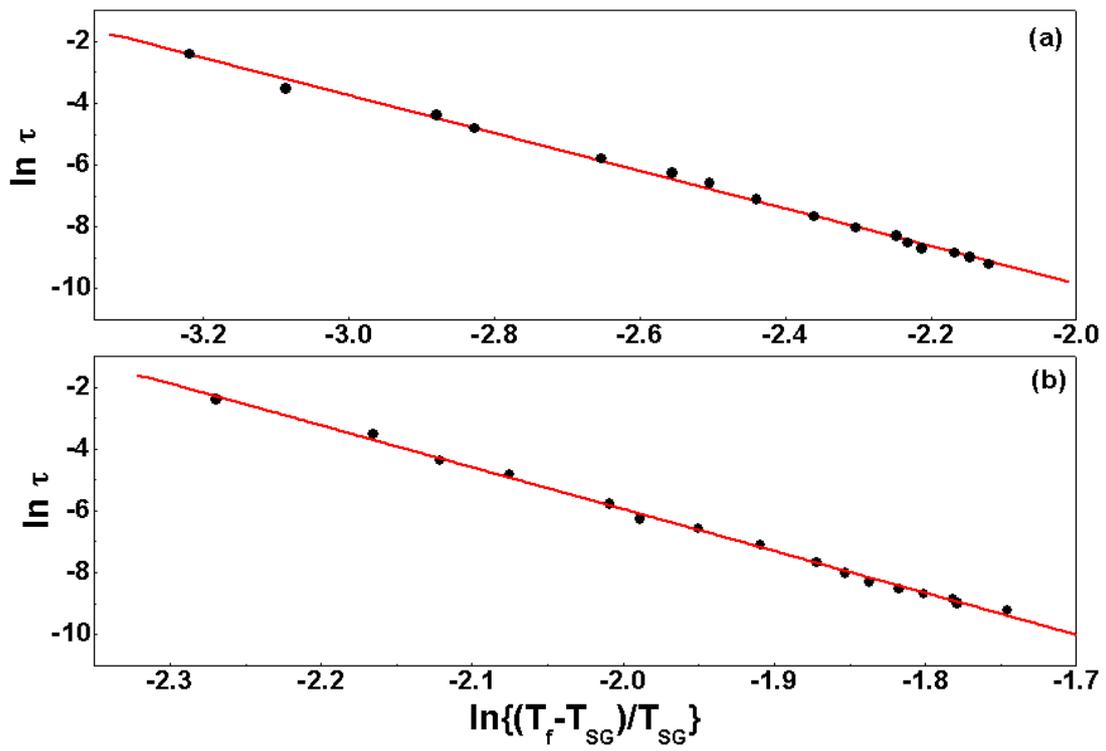

Figure 9

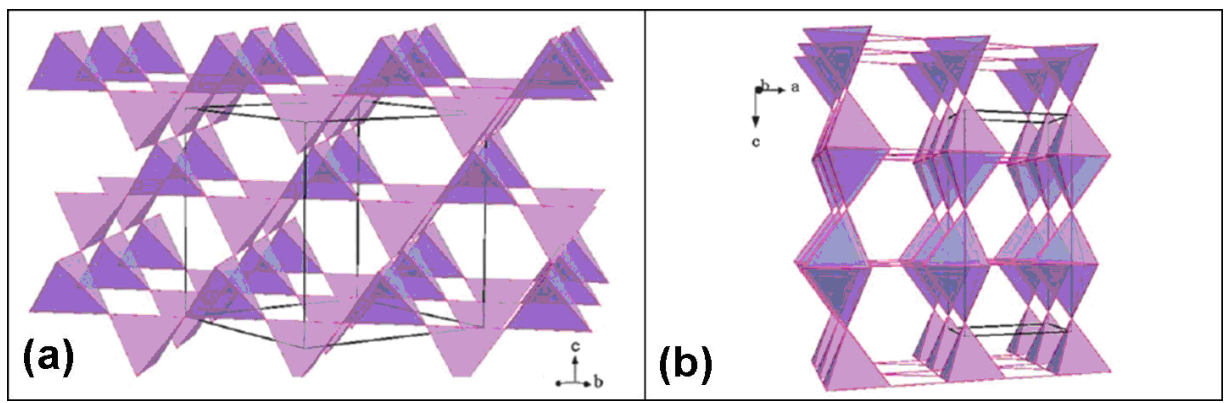

Figure 10